\documentclass[preprint,preprintnumbers,amsmath,amssymb,nofootinbib,superscriptaddress]{revtex4}

\usepackage{graphicx}
\usepackage{dcolumn}
\usepackage{bm}

\begin{document}

\preprint{arXiv:0708.2123}

\title{Multi-quark potential from AdS/QCD}

\author{Wen-Yu Wen}
\email{steve.wen@gmail.com}
\affiliation{Department of Physics and Center for Theoretical Sciences\\ 
National Taiwan University, Taipei 106, Taiwan}


\begin{abstract}
Heavy multi-quark potential in the $SU(N)$ color group
using hard-wall AdS/QCD at both zero and finite temperature is
studied.  A Cornell-like potential is obtained for baryons and other exotic
configurations and compared with those in the quenched lattice
calculation in $N=3$ case.  At the end we also discuss possible
improvements in the UV region of potential.
\end{abstract}

\pacs{11.25.Tq, 74.20.-z}
\keywords{AdS-CFT Correspondence, Heavy Quark Physics}

\maketitle


\section{Introduction}

When the mass of a heavy quark $m_Q$ is much larger than the QCD
scale $\Lambda_{QCD}$, the heavy $\bar{Q}Q$  bound states
(quarkonium) are like hydrogen atoms where ordinary quantum
mechanics can be applied.  In the non-relativistic limit, it is
possible to describe the interaction between $Q$ and $\bar{Q}$ in
terms of a local potential $E(L)$, where $L$ is inter-quark
separation. By solving the Schr\"{o}dinger equation in
three-dimensions, the quarkonium spectrum can be adequately
characterized by $E(L)$ and $m_Q$.  This potential approach to
quarkonia has been applied to the study of Charmonium and
Bottomonium states and a fair agreement with measurement was
found.  One of the most popular models for $E(L)$ is the so-called
{\sl Cornell} potential\cite{Cornell:1978},
\begin{equation}\label{cornell}
E(r)=-\frac{A}{L}+\sigma L.
\end{equation}
The first part is Coulombic due to the one-gluon exchange between
$Q$ and $\bar{Q}$.  The second part is the confinement potential
due to the formation of color flux tube.  Relativistic, radiative
and other non-perturbative corrections can also be added {\sl ad
hoc} as extra terms.  Though we are still lacking a
first-principle derivation of the whole potential, among few
approaches to the non-perturbative QCD, lattice QCD computation
provides an irreplaceable way to test its validity.

On the other hand, the color flux tube can be easily modelled by a
string.  In particular, the coefficient $\sigma$ in the equation
(\ref{cornell}) is identified as the string tension.  Though this
old version of String Theory first failed to be a realistic QCD
model for its false prediction of some massless particles, it
nevertheless took another ambitious task as a candidate of Theory
of Everything and later evolved into its supersymmetric
counterpart, the Superstring Theory.

However, not only QCD string survived but in fact it revived by
gaining new insights from the so-called Anti-de Sitter
space-Conformal Field Theory (AdS/CFT) correspondence, first
proposed in \cite{Maldacena:1997re,Gubser:1998bc,Witten:1998qj}.
In particular, it was applied to relate the thermodynamics of
${\cal N}=4$ super Yang-Mills (SYM) theory in four dimensions to
the thermodynamics of Schwarzschild black holes in
five-dimensional Anti-de Sitter space\cite{Witten:1998zw}.  In
this description, confinement/deconfinement phase transition of
gauge theory on a sphere has its holographic dual description as
the Hawking-Page phase transition\cite{Hawking:1982dh}.  Later, it
was realized that confinement can be achieved by capping off the
Calabi-Yau cone smoothly at the infrared
tip\cite{Klebanov:2000hb,Maldacena:2000yy,Polchinski:2000uf}, or
by introducing IR cutoff in the AdS space
\cite{Polchinski:2001tt,Erlich:2005qh,Andreev:2006eh,Andreev:2006ct,Boschi-Filho:2006pe,Herzog:2006ra,
Ballon Bayona:2007vp,Wen:2007vy}. This latter approach are usually
referred to as the bottom-up construction of AdS/QCD, together
with the other top-down construction by
\cite{Sakai:2004cn,Sakai:2005yt}, are two main approaches to
realize QCD physics via the (super)gravity theory in specific
backgrounds. In particular, it is hoped that, as an alternative to
lattice QCD, the non-perturbative region of QCD can be better
understood both qualitatively and quantitatively within its dual
picture of gravity.  One important testing ground for the AdS/QCD
is the experiments of Relativistic Heavy Ion Collision (RHIC),
collisions of gold nuclei at $200$ GeV per nucleon are about to
produce a strongly-coupled quark-gluon plasma (QGP), which behaves
like a nearly ideal fluid. While the perturbative calculation can
not be fully trusted in this strongly-coupled region, there are
increasing amount of interests in calculation of hydrodynamical
transport quantities via the use of AdS/CFT correspondence, in
particular that the energy loss of a heavy quark moving through
${\cal N}=4$ SYM thermal plasma has been extensively
studied\cite{energy loss}. The drag force was derived in the
context of AdS/CFT to model the effective viscous interaction
\cite{hkkky,gubser}, later it was generalized to a rotating black
hole or with a dilaton field
\cite{herzog,cacg1,cacg2,Nakano:2006js} and
B-field\cite{Matsuo:2006ws,Filev:2007gb}. Drag force of a comoving
$(Q\bar{Q})$ pair was also considered  in \cite{Chernicoff:2006hi}
and energy loss of baryon was studied in \cite{Chernicoff:2006yp}.
It is hoped that this line of research will eventually make
contact with experimental results from RHIC.

Back to the equation (\ref{cornell}), one would like to know how
to derive it in the revived String Theory. We have learnt that
$(Q\bar{Q})$ static potential can be calculated via temporal
Wilson loop for time $T\gg L$,
\begin{equation}
<{\cal W}>\simeq e^{-TE(L)}.
\end{equation}
In the AdS/QCD scenario, this quantity, at zero genus, can be
calculated via on-shell classical (super)gravity action in the AdS
bulk geometry, i.e. the minimal surface enclosed by the Wilson
loop on the boundary\cite{Maldacena:1998im}.  Similarly, baryon
potential can also be constructed once the baryon vertex is
realized as a wrapped brane on the compactified
sphere\cite{Witten:1998xy,Brandhuber:1998xy,Callan:1998iq,Callan:1999zf,Tai:2007gg}.
However, the potential obtained in the AdS background is always
Coulombic thanks to its conformal symmetry.  A Cornell-like
potential for heavy $(Q\bar{Q})$ was obtained in
\cite{Boschi-Filho:2006pe} by breaking this symmetry via new scale
set by an IR cut-off.  This so-called hard-wall model simply puts
an IR cut-off brane at location $r=R$. This cut-off brane is
responsible for confinement as it will become clear later.  The
corresponding metric is then,
\begin{equation}
ds^2=(\frac{r}{R})^2(-dt^2+d\vec{x}^2)+(\frac{R}{r})^2dr^2+R^2d\Omega_5^2
\end{equation}
where $R \leq r < \infty$.

In this paper, we go beyond the study of heavy $(Q\bar{Q})$ static
potential, and look for multi-quark potential in the $SU(N)$ color
group using the above-mentioned toy model of AdS/QCD at both zero
and finite temperature.

This paper is organized as follows.  In section \ref{II} we
refresh the construction of heavy $(Q\bar{Q})$ potential as in
\cite{Boschi-Filho:2006pe}. In section \ref{III} we construct the
heavy baryon potential and make a naive comparison with that in
the quenched lattice calculation for $SU(3)$ color
group\cite{Takahashi:2000te}. In section \ref{IV} we construct
exotic multi-quark configuration. In particular, we obtain
tetra-quark and penta-quark potential and compare them with the
lattice results. In section \ref{V} we study baryon potential at
finite temperature. In section \ref{VI} we make some proposals to
improve our construction, but also discuss their limitations.  In
section \ref{VII}, we conclude with summary and a few comments.

\section{Heavy meson potential}\label{II}
In this section, we recall the construction about inter-quark
potential of heavy $(Q\bar{Q})$ in truncated AdS space.  When the
quark and anti-qaurk are close enough, the potential $E$ and
inter-quark distance $L$ are given by
\begin{eqnarray}
E(r_0)=\frac{r_0}{\pi\alpha'}\bigl(\int_{1}^{\infty}{dy}(\frac{y^2}{\sqrt{y^4-1}}-1)-1\bigr),\nonumber\\
L(r_0)=2\frac{R^2}{r_0}\int_1^{\infty}{dy}\frac{1}{y^2\sqrt{y^4-1}},
\end{eqnarray}
where $r_0$ is the lowest point where QCD string can reach in the
bulk. The $AdS$ ($S^5$) radius
$R^4=\lambda/\alpha'^2=g_{YM}^2N/\alpha'^2$. Therefore one obtains
Coulomb-like potential\cite{Maldacena:1998im}
\begin{equation}
E=-\alpha \frac{\sqrt{\lambda}}{L},\qquad \alpha\simeq 0.228.
\end{equation}
This result is expected from conformal invariance.  As the
distance increases, the U-shape string will be cut by the IR brane
and the potential will include an additional term, given
by\cite{Boschi-Filho:2006pe}
\begin{equation}\label{meson}
E'=E(R)+\frac{1}{2\pi\alpha'}(L-L(R)).
\end{equation}
As a result, we recover linear potential from the second term for
large separation.

\begin{figure}\label{fig1}
\includegraphics[width=0.45\textwidth]{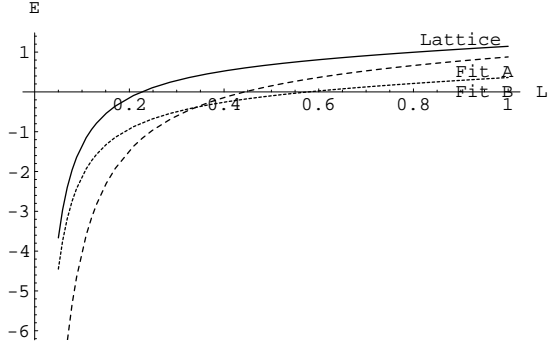}
\caption{Inter-quark potential as a function of
baryon size. Fitting curve A agrees with the lattice simulation at
IR region, while curve B agrees at UV region.}
\end{figure}

\section{Heavy baryon potential}\label{III}
The baryon $(Q\cdots Q)$ in the $SU(N)$ SYM is described by $N$
open strings joining at a vertex formed by wrapping a D5 brane on
$S^5$. The introduction of open strings is essential to
incorporate quarks in fundamental representation of the SYM. The
total energy of this baryon configuration is simply a sum of
energy of $N$ F1's and a wrapped D5 brane, given
by\cite{Brandhuber:1998xy}
\begin{equation}
E=-\beta N \frac{\sqrt{2\lambda}}{L},\qquad \beta \simeq 0.007
\end{equation}
where the vertex is situated at the center of $AdS_5$ and the
total energy has been regularized by subtracting off the energy of
free quarks.

With the IR cut-off, we simply allow the configuration of each
open string to be the same as in the original AdS space as if the
vertex were at $r=r_0<R$. However, those strings are cut by the IR
brane and replaced by string segments on the brane.  Therefore the
vertex is actually projected onto the brane. The relation between
$r_0$ and baryon radius $L$ is still valid, i.e.
\begin{equation}
L=\frac{R^2}{r_0}\int_1^{\infty}\frac{dy}{y^2\sqrt{\beta^2y^4-1}}=
\gamma\frac{R^2}{r_0},\qquad \gamma\simeq 0.481
\end{equation}
and the projected radius $L_R$ on the cut-off is
\begin{equation}
L_R=\frac{R^2}{r_0}\int_1^{y_R}\frac{dy}{y^2\sqrt{\beta^2 y^4-1}},
\end{equation}
where $y\equiv \frac{r}{r_0}$, $y_R\equiv\frac{R}{r_0}$ and
$\beta=\sqrt{16/15}$.  The total energy, composed of one vertex
and $N$ strings, becomes
\begin{equation}
E =
\frac{N}{2\pi\alpha'}r_0\bigl(\int_{y_R}^{\infty}{dy}(\frac{\beta
y^2}{\sqrt{\beta^2y^4-1}}-1)-1\bigr)+\frac{N}{2\pi\alpha'}L_R+\frac{R^5}{8\alpha'}.
\end{equation}
After replacing $r_0$ and $E$ in terms of $L$, and choosing $y_R$
properly, it becomes
\begin{equation}
E = -N\frac{A}{L}+\sigma NL + C.
\end{equation}
In particular, at the limit $r_0\to 0$, one obtains\footnote{$L$
here refers to the projected distance from constituent quark to
the vertex, not the inter-distance between quarks as used in the
lattice simulation\cite{Takahashi:2000te}. If the convention of
lattice is used, we obtain $A'\simeq 0.133\sqrt{\lambda}$, about
half the coefficient of meson.}
\begin{equation}
A=\gamma\frac{\sqrt{\lambda}}{2\pi}, \qquad \sigma =
\frac{1}{2\pi\alpha'},\qquad C=\frac{R^5}{8\alpha'}
\end{equation}
In the FIG. 1, we compare with the quenched lattice
result for $SU(3)$ color group\cite{Takahashi:2000te}, i.e.
$(A,\sigma,C)=(0.0768,0.1524,0.9182)$.  However, two variables are
generally insufficient to fit three unknowns.  If we fit the same
$\sigma$ and $C$ with the choice of $\alpha'=1.0443$ and
$R^2=2.2591$ (or $\lambda=5.1036$), then a twice bigger $A=0.1656$
is obtained for curve A.  If we fit the same $A$ and $\sigma$,
then we obtain $C=0.1345$ (or $\lambda=1.0064$) for curve B. As a
result, fitting curve A agrees with the lattice simulation in the
IR region, while curve B agrees in the UV region.

\begin{figure}\label{fig2}
\includegraphics[width=0.45\textwidth]{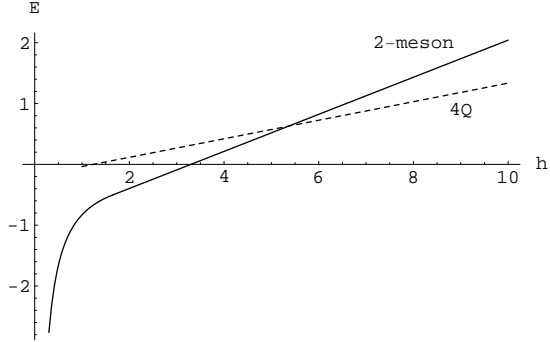}
\caption{Flux-tube recombination
between connected four-quark state and two-meson state is referred
to the {\sl flip-flop}.  It happens as two vertices are close
enough.}
\end{figure}

\begin{figure}\label{fig3}
\includegraphics[width=0.45\textwidth]{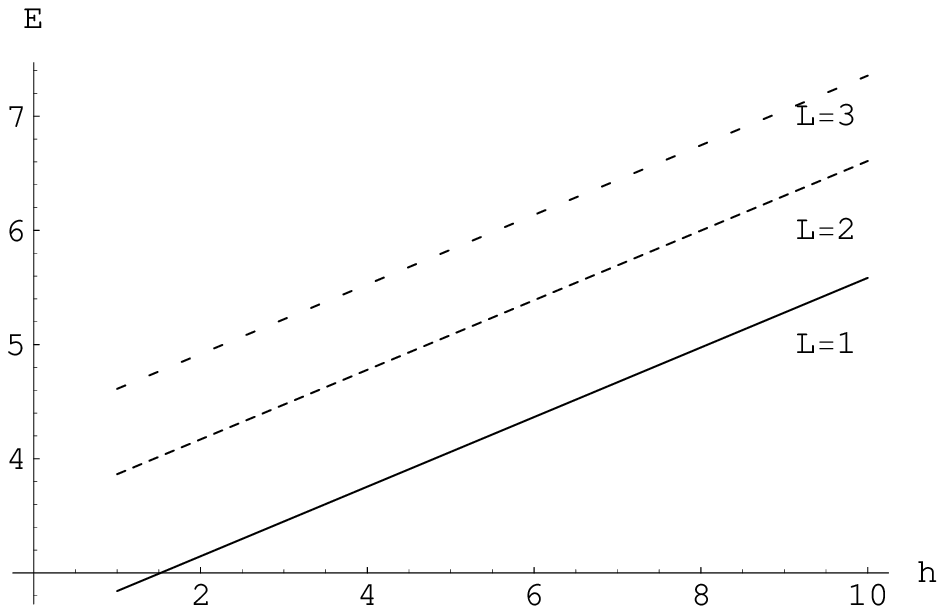}
\includegraphics[width=0.45\textwidth]{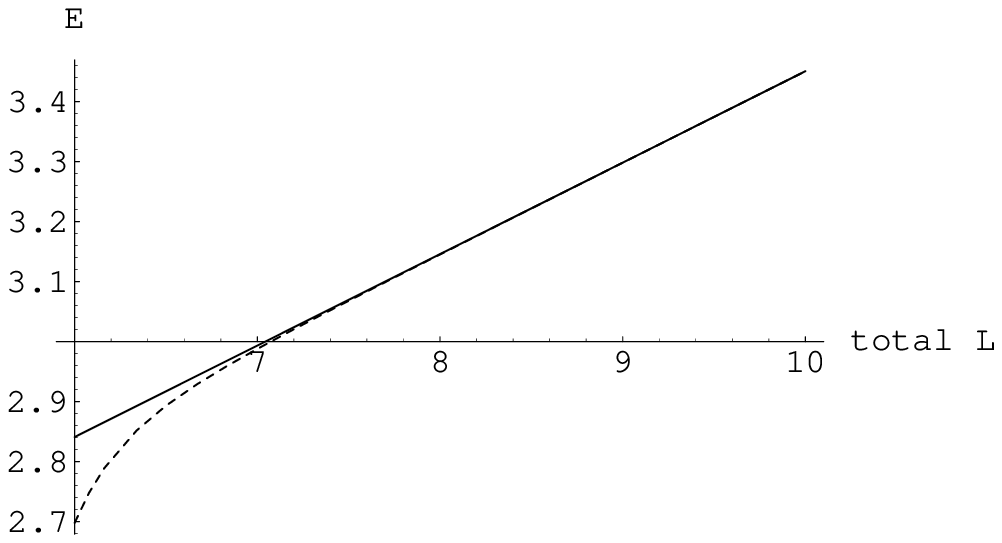}
\caption{To the left: It shows that penta-quark potential is a linear function of vertex separation $h$ for
various quark-vertex distance $L$.  To the right: It shows that penta-quark
potential is a linear function of total flux tube length(solid
line).  The dashed line indicates a deviation from linearity in
the UV region thanks to either the correction from attraction
between vertices or the flip-flop mechanism, as also predicted in
the lattice simulation.}
\end{figure}

\section{Exotic multi-quark potential}\label{IV}
One can imagine that multi-quark system might be formed in the
process of pair annihilation as two or more baryons/anti-baryons
are dragged toward each other.  In the case of $N=3$, for
instance, a baryon made of three quarks $(QQQ)$ can combine with
an anti-baryon made of three anti-quarks $(\bar{Q}\bar{Q}\bar{Q})$
to form a tetra-quark $(QQ\bar{Q}\bar{Q})$, by annihilating a pair
of $(Q\bar{Q})$ and reconnecting two baryon vertices. Similarly, a
penta-quark $(QQ\bar{Q}QQ)$ can be constructed by an anti-baryon
and two baryons. In general, one could form a bound state of
$(mN-2m+2)$ (anti-)quarks with $m$ vertices as long as its energy
is lower than sum of $m$ (anti-)baryons\footnote{Here we simply
mention those bound states made of single open chain of baryon
vertices.  In addition, there might be states with closed
chain(s).  We thank Kazuyuki Furuuchi to point out this.}. From
now on, we restrict ourselves to the case of $N=3$. It is
straightforward to write down inter-quark potential of a
tetra-quark, i.e.
\begin{equation}
E_{4Q}=-4\frac{A}{L}+(4L+h)\sigma +2C,
\end{equation}
where $h$ is the separation between two vertices.  For small $h$,
however, the lattice data tend to agree with the potential of
two-meson, i.e. twice the potential given by (\ref{meson}) with
separation $h$.  This flux-tube recombination between connected
four-quark state and two-meson state is referred to the {\it
flip-flop}. The FIG. 2 shows that this flip-flop can
also be realized in our model.  Next, inter-quark potential of a
penta-quark reads,
\begin{equation}
E_{5Q}=-5\frac{A}{L}+(4L+2h)\sigma + 3C,
\end{equation}
where we have assumed that three vertices are placed in equal
separation $h$. From the equation above, it is obvious to expect
linear relation between potential $E$ and separation $h$ as shown
in the left plot of FIG. 3, or between $E$ and total length of
color flux tube, say $4L+2h$, as shown in the right plot of FIG. 3.

\begin{figure}\label{fig4}
\includegraphics[width=0.45\textwidth]{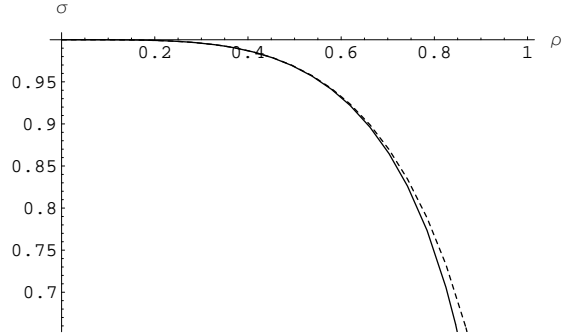}
\caption{The solid curve shows the
temperature dependence of string tension (normalized to unity). It
can be very well fit by the dashed curve given by equation
(\ref{fitT}) for low temperature.}
\end{figure}

\section{Potential at finite temperature}\label{V}
Now we are ready to study the same potential at the finite
temperature, which corresponds to introducing an AdS black hole,
i.e.
\begin{equation}
ds^2=(\frac{r}{R})^2(-f(r)dt^2+d\vec{x}^2)+(\frac{R}{r})^2\frac{1}{f(r)}dr^2+R^2d\Omega_5^2,
\end{equation}
where $f(r)=1-r_T^4/r^4$ and the event horizon is related to the
Hawking temperature by $r_T=\pi R^2 T$.  In the deconfined phase
where $r_T\ge R$, the vertex falls into the black hole and leaves
behind $N$ dissociated quarks with total mass\footnote{This bare
mass is not physical but by convention of renormalization since
ideal heavy quarks should have infinite mass which is
unmeasurable.},
\begin{equation}
E=-N\frac{r_T}{2\pi\alpha'}.
\end{equation}
However, the observers living in the boundary still see the vertex
since it takes infinite time for the vertex to cross the horizon.
In the confining phase where $r_T < R$, one may either argue that
black hole is unaccessible to us so the potential has no
temperature dependence for thermal AdS, or one may pretend that
black hole still curves the bulk geometry.  In the latter
situation, the baryon potential has the following linear
component\cite{Brandhuber:1998xy},
\begin{equation}
N\frac{\sqrt{\lambda}}{2\pi
r_0}\int_1^{y_R}{dy}\sqrt{\frac{15-18\rho^4-\rho^8}{(y^4-\rho^4)(16y^4-15+2\rho^4+\rho^8)}},
\end{equation}
where $\rho=r_T/r_0$.  Here, different from the situation at zero
temperature, it is not allowed to send $r_0\to 0$.  In fact, the
reality condition requires that $r_0>r_T$.  In the FIG. 4 we see the temperature dependence of the effective
string tension, which can be best fit in low temperature by
\begin{equation}\label{fitT}
\frac{\sigma(T)}{\sigma(0)} \simeq \sqrt{1-\rho^4}.
\end{equation}

\section{Improvements}\label{VI}
We have argued in section \ref{III} that our model can not
completely fit the lattice result.  In this section, we provide
three possible improvements to the potential but also discuss
their limitations.

\subsection{UV perfection}
As mentioned before that our model cannot completely fit the
lattice result, introducing another UV cut-off $r_c$ may resolve
this.  However, this implies that instead of our original
assumption of (infinitely) heavy quarks, the Cornell potential is
realized via quarks with a finite mass
\begin{equation}
m_Q \simeq \frac{r_c-R}{2\pi\alpha'}.
\end{equation}
If this is the case, we should also consider string breaking (pair
production) as inter-quark separation grows. A careful examination
excludes the possibility of sending $r_0\to 0$, which leaves more
than one choice for combination of $r_0$ and $r_c$, or
equivalently $y_R$ and $y_c=r_c/r_0$.  In fact, we only find
reasonable solutions as $y_R<2$.  One of these choices is that,
for instant, $y_R\sim1.50$ and $y_c\sim3.29$.

\subsection{IR refinement}
We start with a truncated model with IR cut-off at $r=R$. This
implies that the confinement/deconfinement transition temperature
is at $T_c=1/\pi R$, which is $\simeq 212$ MeV for the first fit A
and $\simeq 312$ MeV for the second fit B.  However, this model
may be refined with a different IR cut-off $r_{IR}$.  This
provides a new parameter for perfect fit, where
\begin{equation}
C=\frac{R^4}{\alpha'}r_{IR}.
\end{equation}
with the choice of $r_{IR}\simeq 6.9891$, However, this cut-off
corresponds to the transition temperature\cite{Herzog:2006ra},
\begin{equation}
T_c = \frac{2^{1/4}}{\pi r_{IR}}\simeq 54\quad MeV,
\end{equation}
which is, unfortunately, far too low in comparison with the
lattice prediction $T_c\simeq 190$ MeV\cite{Karsch:2006xs}.

\subsection{Correction via closed string channel and flip-flop mechanism}
The lattice simulation indicates that in the FIG. 3,
relation between potential and length is not completely linear,
especially for small separation $h$.  The residual force outside
each interacting group of quarks may help explaining the deviation
from linearity. This could be realized in our model by two means:
one way is to include the graviton (closed string) exchange
between vertices.  The other is the flip-flop mechanism in short
$h$.  Both give the expected $1/h$ behaviour in the IR region.

\section{Conclusion}\label{VII}
In this paper, we have constructed static potential of heavy
baryon and other exotic multi-quark configurations in the
hard-wall model of AdS/QCD.  In particular, we obtained a
Cornell-like potential and make naive comparison with that in the
quenched lattice calculation for $SU(3)$ color
group\cite{Takahashi:2000te}.  We found that the flux tube
recombination between four-quark and two-meson states (flip-flop)
happens when two vertex approaches each other in a tetra-quark
configuration.  In a penta-quark configuration, static potential
is always linear with respect to inter-vertex distance or total
length of flux tube, though a deviation from linearity is expected
for short $h$. We later proposed three different ways to improve
our model.  For UV perfection, we introduced a UV cut-off and
there were more than one choices to achieve the goal.
Nevertheless, the quark mass became finite and one has to take
into account pair production (string breaking) as inter-quark
distance grows. For IR refinement, we traded a tunable IR cut-off
for perfect fit. However, this modification gave us far too low
confinement/deconfinement transition temperature and was not so
impressive.  Finally, we argued that the deviation from linearity
in the UV region in the figure (\ref{fig3}) may come from
correction via vertex interaction or flip-flop mechanism when the
potential of two-meson state dominates.  At end end, we would like
to make a few remarks. First of all, there is no obvious reason to
believe that AdS/QCD construction is trusted for large $N$ as well
as in the AdS/CFT correspondence.  On the top of that, we should
not ask too much for quantitative accuracy even for small $N$,
especially without taking in account the $1/N$ correction. This
may well explain that we did not find a perfect fit without
causing another drawback. Secondly,  in the confined phase, the
string tension is expected to have a temperature correction
$\propto -T^2$ instead of equation (\ref{fitT}), observed by
\cite{Boschi-Filho:2006pe} as well. It is interesting to see if we
could settle this disagreement in the context of AdS/QCD. Thirdly,
there was early debate on $Y$-shape or $\Delta$-shape flux tube
configuration in the baryon ground state. Our construction always
prefers the former since the latter costs more energy for slightly
longer tube length and two extra vertex.  At last, our proposal of
UV perfection has similar effect when flavored quark is
introduced\footnote{The author would like to thank Carlos Nunez
for pointing out this similarity.}, and a warping geometry with
linear dilaton is responsible to the string
breaking\cite{Casero:2006pt}. It would be interesting to realize
this breaking scenario in the soft-wall QCD model.  In conclusion,
the new feature as well as advantage of this construction, in
comparison with the old version QCD string, is to realize all
parts of Cornell potential, though not perfect, by one piece of
configuration of strings and vertices stretching over truncated
AdS bulk.  This implies that there may be a chance in AdS/QCD
construction for a unified theory of QCD in both perturbative and
non-perturbative regions.\\


\begin{acknowledgments}
I am grateful to Kazuyuki Furuuchi, Pei-Ming Ho,
Namit Mahajan, and Carlos Nunez for useful discussion. I wish to
thank Kazuyuki Furuuchi for a careful reading of the manuscript.
The author is supported in part by the Taiwan's National Science
Council under grant NSC96-2811-M-002-018.
\end{acknowledgments}

\bibliography{apssamp}

\end{document}